# بهبودی بر روش‌های درج در کم‌ارزش‌ترین بیت تطبیقی و درج در کم‌ارزش‌ترین بیت تطبیقی بازبینی شده

کاظم غضنفری[1]، رضا صفابخش[2]

[1]دانشکده مهندسی کامپیوتر، دانشگاه صنعتی امیرکبیر، تهران، Kazemmit@aut.ac.ir

[2]دانشکده مهندسی کامپیوتر، دانشگاه صنعتی امیرکبیر، تهران، Safa@aut.ac.ir

چکیده – هدف از نهان‌نگاری ایجاد یک کانال محرمانه و امن بین فرستنده و گیرنده است. دو عدد از رایج‌ترین و ساده‌ترین روش‌های ایجاد این کانال، روش‌های درج در کم‌ارزش‌ترین بیت تطبیقی و درج در کم‌ارزش‌ترین بیت تطبیقی بازبینی شده می‌باشد. امن بودن در مقابل اکثر حملات آماری مرتبه اول مهمترین ویژگی این دو روش می‌باشد. از طرف دیگر، این دو روش همبستگی موجود بین پیکسل‌ها را در نظر نمی‌گیرند، لذا، اخیراً ماتریس خودهمبستگی که حاوی اطلاعات همبستگی موجود بین پیکسل‌ها است، توسط تعداد زیادی از روش‌های نهان‌کاوی برای تشخیص تصاویر نهان‌نگاری شده توسط این دو روش، مورد استفاده قرار گرفته است. بنابراین اگر بتوان عملیات کاهش یا افزایش مقدار پیکسل‌های تصویر توسط این روش‌ها را تطبیقی نمود بصورتیکه منجر به میزان تغییرات کمتری روی ماتریس‌های خودهمبستگی شود، این دو روش به مقدار زیادی در مقابل روش‌هایی که از این اطلاعات آماری استفاده می‌کنند امن خواهند شد. در این مقاله بهبود مذکور بر روی روش‌های درج در کم‌ارزش‌ترین بیت تطبیقی و درج در کم‌ارزش‌ترین بیت تطبیقی بازبینی شده اعمال خواهد شد. نتایج آزمایشات نشان می‌دهد که بهبود پیشنهادی منجر شده است که تغییرات بوجود در ماتریس خودهمبستگی بسیار کاهش یافته و این دو روش در مقابل حملاتی که از ماتریس خودهمبستگی استفاده می‌کنند امن شوند.

کلمات کلیدی- نهان‌نگاری، نهان‌کاوی، درج در کم ارزش‌ترین بیت تطبیقی، درج در کم‌ارزش‌ترین بیت تطبیقی بازبینی شده، ماتریس خودهمبستگی

## 1- مقدمه

با توجه به رشد سریع ارتباطات و اینترنت، انتقال داده‌ها در قالب دیجیتال امری مرسوم شده است. یک مسئله مهم در رابطه با انتقال اطلاعات دیجیتال از طریق اینترنت آن است که امنیت انتقال اطلاعات تضمین شود. حال هدف از نهان‌نگاری، ارسال پیام‌های محرمانه بصورت مخفیانه روی شبکه می‌باشد، بطوریکه مهاجمان متوجه انتقال داده‌های مخفی و امنیتی در کانال ارتباطی نشوند[1،2].

روش‌های گوناگونی برای نهان‌نگاری اطلاعات در تصویر ارائه شده است. در یک دسته‌بندی، روش‌های نهان‌نگاری را بر اساس حوزه درج به دو دسته اصلی تقسیم بندی می‌کنند: نهان‌نگاری در حوزه مکان و نهان‌نگاری در حوزه تبدیل. روش‌های نهان‌نگاری در حوزه مکان، اطلاعات محرمانه را بطور مستقیم در مقدار پیکسل‌ها درج می‌کنند، درحالیکه روش‌های نهان‌نگاری در حوزه تبدیل ابتدا تصویر را به کمک یکی از تبدیلات مانند $DCT$، $DWT$ و $DFT$ به حوزه فرکانس انتقال می‌دهند، سپس عمل نهان‌نگاری را در این حوزه انجام می‌دهند. درج اطلاعات در تصویر منجر به تغییرات گوناگونی می‌شود که به سه دسته کلی تقسیم می‌شوند: تغییرات ادراکی، آماری و ساختاری، که این تغییرات زمینه تشخیص وجود پیام محرمانه در تصویر را برای مهاجمان فراهم می‌نماید[1،2].

یکی از مهم‌ترین اطلاعات آماری مراتب بالا ماتریس خودهمبستگی می‌باشد. بر اساس اینکه عمل نهان‌نگاری در چه حوزه‌ای انجام شده است، این ماتریس قابل محاسبه می‌باشد و ما نیازمند روش‌هایی هستیم که این ماتریس را برای تصویر (یا ضرایب تبدیل) پس از درج در حوزه مکان (یا تبدیل) حفظ کنند.

دو عدد از رایج‌ترین روش‌های نهان‌نگاری درج در کم‌ارزش‌ترین بیت تطبیقی و درج در کم‌ارزش‌ترین بیت تطبیقی بازبینی شده می‌باشد. مهمترین ویژگی این روش‌ها آن است که در مقابل حملاتی که از اطلاعات آماری مرتبه اول استفاده می‌کنند امن می‌باشند. از طرف دیگر، اخیراً روش‌های نهان‌کاوی ارائه شده است که با بهره بردن از ماتریس خودهمبستگی اقدام به تشخیص تصاویر نهان‌نگاری شده توسط این دو روش می‌کنند [3-6].

در این مقاله با اعمال بهبودی به دو روش نهان‌نگاری مذکور سعی بر آن است که این دو روش را در مقابل این حملات امن کنیم. در

بخش ۲، اطلاعات آماری مراتب بالا بخصوص ماتریس خودهمبستگی و تاثیر عملیات درج روی آن شرح داده و تعدادی از روش‌های نهان‌کاوی که از این ماتریس بهره برده‌اند را شرح می‌دهیم. در بخش ۳ بصورت مختصر روش‌های نهان‌نگاری درج در کم‌ارزش‌ترین بیت تطبیقی و درج در کم‌ارزش‌ترین بیت تطبیقی بازبینی شده را شرح خواهیم داد. بهبود اعمال شده در بخش ۴ ارائه خواهد شد. در نهایت بهبود اعمال شده در بخش ۵ مورد ارزیابی قرار خواهد گرفت.

## ۲- اطلاعات آماری مراتب بالا

بطور کلی اطلاعات آماری قابل استخراج از تصویر را به دو دسته اصلی تقسیم می‌کنند که شامل اطلاعات آماری مرتبه اول و اطلاعات آماری مراتب بالاتر می‌باشد. یکی از مهم‌ترین اطلاعات آماری مرتبه اول، هیستوگرام تصویر می‌باشد. با توجه به اینکه انجام عمل نهان‌نگاری باعث بوجود آمدن تغییرات قابل توجهی در هیستوگرام تصویر می‌شود، روش‌های زیادی از این تغییرات استفاده کرده و اقدام به نهان‌کاوی کرده‌اند[۷,۸]. برای مقابله با این حمله روش‌های زیادی ارائه شده است که می‌توان به روش‌های درج در کم‌ارزش‌ترین بیت تطبیقی و تکنیک درج در کم‌ارزش‌ترین بیت تطبیقی بازبینی شده اشاره نمود.

هیستوگرام تصویر وابستگی بین پیکسل‌ها را در نظر نمی‌گیرد، لذا این ویژگی جزء اطلاعات آماری مرتبه اول در نظر گرفته می‌شود. اطلاعات آماری مراتب بالاتر، به اطلاعاتی گفته می‌شود که وابستگی پیکسل‌ها را نسبت به یکدیگر در نظر گیرد. یکی از مهم‌ترین اطلاعات آماری مراتب بالا که در نهان‌کاوی مورد استفاده قرار می‌گیرد، ماتریس خودهمبستگی می‌باشد. به دلیل اینکه انجام عمل نهان‌نگاری باعث می‌شود این اطلاعات تغییر پیدا کند، روش‌های زیادی از این ماتریس بهره گرفته و اقدام به نهان‌کاوی نموده‌اند [۳-۶].

یک روش برای توصیف همبستگی بین پیکسل‌ها استفاده از ماتریس‌های خودهمبستگی است. ماتریس خودهمبستگی، یک ماتریس دو بعدی است که مدخل‌های آن بیانگر فرکانس رخ دادن دو سطح خاکستری است که به فاصله و در جهت خاصی قرار دارند. زوج ($\Delta x, \Delta y$) بیانگر فاصله و جهت می‌باشند. این ماتریس به نوعی بیانگر اطلاعات آماری مرتبه دو می‌باشد، چون همبستگی پیکسل‌ها را دوبه‌دو در نظر گرفته است. برای سطح خاکستری $g(x,y)$ مربوط به پیکسل در مکان $(x,y)$ مدخل $c_{i,j}$ ماتریس خودهمبستگی، فرکانس تعداد زوج‌هایی را بیان می‌کند که در رابطه (۱) صدق کنند.

$$(g(x,y) = i) \wedge (g(x+\Delta x, y+\Delta y) = j) \quad (1)$$

در اینصورت برای هر زوج ($\Delta x, \Delta y$) یک عدد ماتریس خودهمبستگی ایجاد می‌شود. چون همبستگی موجود بین پیکسل‌های مجاور نسبت به همبستگی موجود بین پیکسل‌هایی که دارای فاصله زیادی از هم می‌باشند، بسیار بیشتر است. به همین دلیل در روش‌های نهان‌کاوی معمولاً ماتریس‌های خودهمبستگی را فقط برای پیکسل‌های همسایه محاسبه می‌کنند. به عبارت دیگر، ماتریس‌های همبستگی را به ازای زوج‌های زیر محاسبه می‌کنند.

$(\Delta x, \Delta y) \in \{(1,0), (-1,1), (0,1), (1,1),$
$(-1,-1), (0,-1), (1,-1), (-1,0)\}$

البته در روش‌های گوناگون، تمام همسایگی‌های فوق مورد استفاده قرار نگرفته است. بطور مثال ابوالقسمی [۳] و جوون [۴] زوج‌های
$(\Delta x, \Delta y) \in \{(-1,1)(0,1)(-1,-1)(-1,0)\}$
، سان و همکارانش [۵] زوج‌های
$(\Delta x, \Delta y) \in \{(1,0)(0,1)\}$
و فرنز [۶] زوج‌های زیر مورد استفاده قرار گرفته است:
$(\Delta x, \Delta y) \in \{(1,0)(-1,1)(0,1)(1,1)\}$

با توجه به اینکه در تصاویر سطح خاکستری برای نمایش مقدار پیکسل‌ها از ۸ بیت استفاده می‌شود، ابعاد این ماتریس برابر ۲۵۶×۲۵۶ خواهد بود. البته سان و همکارانش [۵] برای کاهش حجم محاسبات و ابعاد ماتریس ابتدا مشتق تصویر محاسبه شده، بعد از یک مرحله آستانه‌سازی (شبه آستانه‌سازی)مقادیر تصاویر مشتق، اقدام به محاسبه این ماتریس برای تصویر مشتق آسانه‌ای شده کرده است. واضح است در اینصورت ابعاد ماتریس به شدت کاهش پیدا می‌کند.

چون در ماتریس خودهمبستگی ارتباط پیکسل‌ها دوبه‌دو در نظر گرفته می‌شود، توصیف تاثیر درج روی این ماتریس بسیار سخت است. در تصاویر طبیعی سطح خاکستری پیکسل‌های مجاور پیوسته و نرم می‌باشد. از طرف دیگر چون این ماتریس را فقط برای همسایه‌های پیکسل محاسبه می‌کنند، بیشترین انرژی موجود در ماتریس خودهمبستگی روی قطر اصلی و عناصر نزدیک به قطر اصلی این ماتریس تمرکز یافته است. حال عملیات درج را در نظر بگیرید. چون عملیات درج باعث می‌شود که پیوستگی موجود بین مقادیر پیکسل‌های همسایه کاهش یابد لذا انرژی متمرکز شده روی قطر اصلی و عناصر نزدیک به آن کاهش یافته و این انرژی بین عناصر دورتر از قطر اصلی پخش می‌شود. روش‌های نهان‌کاوی مذکور از این تغییرات برای تشخیص تصاویر نهان‌نگاری شده بهره می‌برند.

## ۳- تکنیک‌های درج در کم‌ارزش‌ترین بیت تطبیقی و بازبینی شده آن

در این بخش دو روش نهان‌نگاری رایج که قصد بهبود آنها را داریم بصورت خلاصه شرح می‌دهیم. این دو روش درج در کم‌ارزش‌ترین بیت تطبیقی و درج در کم‌ارزش‌ترین بیت تطبیقی بازبینی شده می‌باشد. وجه اشتراک این دو روش آن است که در عملیات درج می‌توان مقدار یک پیکسل را بطور دلخواه کاهش یا افزایش داد. در ادامه جزئیات این دو روش بیان می‌شود.

## 3-1- درج در کم‌ارزش‌ترین بیت تطبیقی

سادگی درج بروش کم‌ارزش‌ترین بیت تطبیقی همانند روش درج در کم‌ارزش‌ترین بیت ساده می‌باشد. برای این منظور در صورتیکه مقدار بیت پیام که قصد درج نمودن آن را داریم با کم‌ارزش‌ترین بیت مقدار پیکسل برابر باشد، تغییری در مقدار پیکسل داده نمی‌شود. ولی اگر بیت پیام با کم‌ارزش‌ترین بیت مقدار پیکسل برابر نباشد به احتمال برابر مقدار پیکسل مربوطه کاهش یا افزایش می‌یابد. واضح است در صورت کاهش و یا افزایش مقدار پیکسل، کم‌ارزش‌ترین بیت آن برابر بیت پیام خواهد شد. رابطه (۲) بیانگر نحوه درج پیام به این روش می‌باشد. در این رابطه $I_C$ و $I_S$ به ترتیب مقدار یک پیکسل قبل و بعد از درج بیت پیام b می‌باشد. همچنین در این رابطه L برابر ۸ می‌باشد.

$$I_S = \begin{cases} 1 & b \neq LSB(I_C) \,\&\, I_C = 0 \\ I_C \pm 1 & b \neq LSB(I_C) \,\&\, 0 < I_C < 2^L \\ I_C & b = LSB(I_C) \\ 2^L - 2 & b \neq LSB(I_C) \,\&\, I_C = 2^L - 1 \end{cases} \quad (2)$$

همانطور که از رابطه (۲) مشخص است، برای ۵۰ درصد بیت‌های پیام مقدار پیکسل می‌تواند بطور دلخواه کاهش یا افزایش یابد.

## 3-2- درج در کم‌ارزش‌ترین بیت تطبیقی بازبینی شده

در این روش جهت کد کردن ۲ بیت بطور همزمان از ۲ پیکسل استفاده می‌شود و عمل کد کردن این ۲ بیت به این صورت است که حداکثر یکی از دو پیکسل نیاز به تغییر دارند[۹]. اگر $S1, S2$ دو بیت رمز و $y_1, y_2$ محتوای دو پیکسل باشند، تابع باینری LSB به صورت رابطه (۳) تعریف می‌شود.

$$f(y_1, y_2) = LSB\left(\left[\frac{y_1}{2}\right] + y_2\right), \quad (3)$$

که در آن $LSB(x)$ کم‌ارزش‌ترین بیت عدد x را برگشت می‌دهد. فلوچارت شکل ۱ نحوه درج کردن پیام را نمایش می‌دهد که در آن $\hat{y_1}$ و $\hat{y_2}$ داده‌هایی اند که جایگزین $y_1$ و $y_2$ می‌شود. جهت استخراج پیام از رابطه (۴) استفاده می‌شود.

$$s_1 = LSB(\hat{y_1}) \quad , \quad s_2 = f(\hat{y_1}, \hat{y_2}) \quad (4)$$

شکل ۱ فلوچارت درج روش کم‌ارزش‌ترین بیت تطبیقی بازبینی شده را نمایش می‌دهد. همانطور که در این فلوچارت می‌بینید، در یکی از برگ‌های آن (که دارای کادر پررنگ است) به گیرنده اجازه داده شده است که مقدار پیکسل را کاهش یا افزایش دهد. احتمال اینکه در محاسبات روش کم‌ارزش‌ترین بیت تطبیقی بازبینی شده به برگ مذکور برسیم برابر ۰/۲۵ می‌باشد. بهبودی را که ما سعی داریم در این مقاله اعمال کنیم، شامل این برگ (۲۵ درصد عملیات درج) می‌شود.

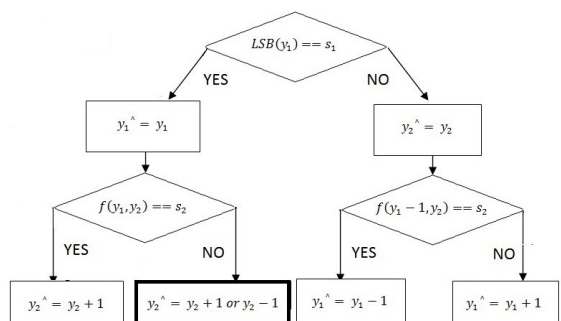

شکل ۱- فلوچارت تکنیک درج در کم‌ارزش‌ترین بیت تطبیقی

## 4- بهبود اعمال شده

همانطور که بیان شد هدف بهبودهای اعمال شده آن است که عملیات درج منجر به تغییرات کمتری در ماتریس خودهمبستگی شود. مسلماً هر چه میزان این تغییرات کمتر شود، روش نهان‌نگاری مربوطه بیشتر در مقابل حملاتی که از این ماتریس استفاده می‌کنند امن خواهد شد. ماتریس خودهمبستگی حاوی اطلاعات مفیدی در رابطه همبستگی مقدار پیکسل‌ها است. در واقع اگر عملیات درج باعث شود مقدار یک پیکسل نسبت به مقادیر پیکسل‌های همسایه آن بیشتر تغییر کند، (همبستگی موجود بین پیکسل‌ها بیشتر تحت تاثیر قرار گیرد) اطلاعات آماری مراتب بالا بیشتر تغییر خواهد کرد. بنابراین اگر بتوان عملیات درج را بصورتی انجام داد که اختلاف مقدار پیکسل نسبت به همسایگانش تا حد امکان کم باشد، ماتریس خودهمبستگی کمتر دچار تغییرات می‌شود.

با توجه به اینکه در دو روش نهان‌نگاری که قصد بهبود آنها را داریم، کاهش و یا افزایش مقدار یک پیکسل (در ۵۰٪ موارد برای روش درج در کم‌ارزش‌ترین بیت تطبیقی و در ۲۵٪ موارد برای روش درج در کم‌ارزش‌ترین بیت تطبیقی بازبینی شده) به دلخواه انجام می‌شود، ما می‌خواهیم عملیات تغییر (کاهش یا افزایش) را به گونه انجام دهیم که اختلاف مقدار پیکسل مربوطه با همسایگانش حداقل باشد. با توجه به اینکه برای محاسبه ماتریس خودهمبستگی غالباً از ۸ همسایه هر پیکسل استفاده می‌نمایند، بنابراین ما هم سعی می‌کنیم با در نظر گرفته مقدار ۸ پیکسل همسایه پیام محرمانه را درج نماییم. برای این منظور کافی است اختلاف بوجود آمده ناشی از کاهش و افزایش را محاسبه کرده و در نهایت حالتی (کاهش یا افزایش) را که منجر به تغییرات کمتر می‌شود اعمال کنیم.

لازم به ذکر است که هر چه اختلاف بین دو مقدار پیکسل همسایه بیشتر باشد، تغییر یکی از آنها تاثیر کمتری روی ماتریس خودهمبستگی دارد. همچنین روش‌های نهان‌کاوی که از ماتریس خودهمبستگی استفاده نموده‌اند فقط از عناصر موجود روی قطر اصلی و عناصر نزدیک به قطر اصلی بهره برده‌اند. شکل ۲ نشان‌دهنده این عناصر از ماتریس خودهمبستگی است. در این شکل عناصری که به

رنگ خاکستری نشان داده شده‌اند بیشتر در روش‌های نهان‌کاوی مورد استفاده قرار گرفته‌اند.

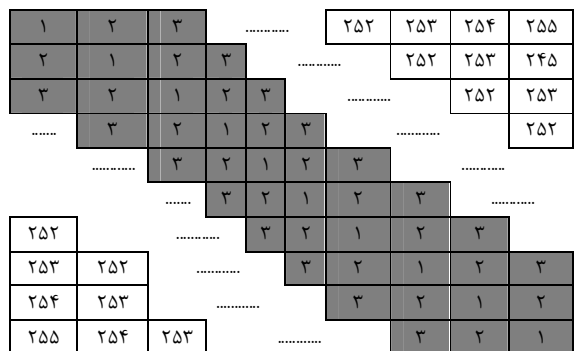

**شکل ۲- عناصری که با رنگ خاکستری نمایش داده شده‌اند، در روش‌های نهان‌کاوی استفاده شده‌اند.**

```
Matrix=double (Cover_Image(i-1:i+1, j-1:j+1));
Matrix_temp=Matrix-Matrix(2,2);
Mask=Matrix_temp <T;

Matrix_P=Matrix;
Matrix_N=Matrix;

Matrix_P(2,2)=Matrix(2,2)-1;
Matrix_N(2,2)=Matrix(2,2)+1;

Matrix_P_diff=(Matrix_P - Matrix_P(2,2)) .* Mask
Matrix_P_diff_SUM=SUM (Matrix_P_diff(:));

Matrix_N_diff=( Matrix_N - Matrix_N(2,2)) .* Mask;
Matrix_N_diff_SUM =SUM (Matrix_N_diff(:));

If  Matrix_P_diff_SUM >Matrix_M_diff_SUM
    Cover_Image(i-1:i+1, j-1:j+1)=Matrix_N(2,2);
else
    Cover_Image(i-1:i+1, j-1:j+1)=Matrix_P(2,2);
End
```

**شکل ۳- کد نحوه محاسبه ماتریس ماسک و نحوه درج پیام در روش بهبود یافته.**

بنابراین همسایگان یک پیکسل، فقط در صورتی روی مقدار آن پیکسل (کاهش و یا افزایش به مقدار یک) تاثیر دارند که اختلاف اولیه آنها از آن پیکسل از حد معینی کمتر باشد. فرض کنید $B$ ماتریسی باشد که $B(2,2)$ پیکسل مورد نظر برای درج و سایر عناصر همسایگان آن پیکسل باشند. در اینصورت ماتریس ماسک (Mask) را بصورت زیر تعریف می‌کنیم. این ماتریس بیانگر آن است که کدام عناصر در تصمیم‌گیری برای تعیین کاهش و یا افزایش مقدار پیکسل موثر می‌باشند. به عبارت دیگر اگر $Mask(i,j)$ برابر ۱ باشد، آن یک عنصر موثر محسوب خواهد شد. رابطه (۵) بیانگر این مطلب است.

بعد از اینکه ماتریس فوق محاسبه شد، کافی است میزان اختلاف بوجود آمده ناشی از کاهش و افزایش پیکسل مورد نظر محاسبه شود. در اینصورت هر کدام از تغییرات مذکور (کاهش، افزایش) که منجر به تغییرات کمتری شود برای فرآیند درج انتخاب می‌شود. کد شکل ۳ با زبان برنامه‌نویسی MATLAB بیانگر این مطب است. توجه شود این کد برای حالاتی از درج توسط دو روش نهان‌نگاری بیان شده می‌باشد که امکان کاهش یا افزایش مقدار پیکسل بطور دلخواه فراهم شده است.

$$\text{Mask}(i,j)_{\substack{i,j \in (1,2,3) \\ (i,j) \neq (2,2)}} = |B(2,2) - B(i,j)| < T \qquad (5)$$

## ۵- آزمایشات

برای انجام آزمایشات از پایگاه‌های تصویری UCID [۱۰] و VASC [۱۱] استفاده شد که مجموعاً حاوی ۵۰۰۰ تصویر بودند. این تصاویر را به تصاویر سطح خاکستری با ابعاد ۵۰۰×۵۰۰ تبدیل کردیم. همچنین عبارات LSB±1 و LSB±1(R) به ترتیب به تکنیک درج در کم‌ارزش‌ترین بیت تطبیقی و بازبینی شده آن اشاره دارد.

همانطور که قبلاً شرح دادیم در تصاویر طبیعی بیشترین مقدار انرژی در قطر اصلی و عناصر نزدیک به قطر اصلی در ماتریس خودهمبستگی تمرکز یافته است و عملیات درج توسط این دو روش باعث می‌شود انرژی این نواحی کاهش یافته و از قطر اصلی بیشتر فاصله بگیرد. در آزمایش اول قصد داریم روش‌های بهبود یافته را به لحاظ میزان تغییرات بوجود آمده در ماتریس خودهمبستگی مورد ارزیابی قرار دهیم. برای این منظور ما پیام به طول 0.4bpp و 0.8bpp را در تصاویر مورد استفاده درج کردیم. شکل ۴ میزان انرژی قطر اصلی و قطرهای نزدیک به این قطر را برای روش درج در کم‌ارزش‌ترین بیت تطبیقی و درج در کم‌ارزش‌ترین بیت بازبینی شده نمایش می‌دهند.

همچنین ما روش‌های بهبود یافته خود را به لحاظ میزان امن بودن در مقابل تعدادی از روش‌های نهان‌کاوی که از ماتریس خودهمبستگی بهره برده‌اند مورد ارزیابی قرار دادیم. جدول ۱ نتایج بدست آمده از این آزمایش را برای حالتی که پارامتر $T$ برابر ۴ باشد نمایش می‌دهد. همانطور که این جدول نمایش می‌دهد بهبود اعمال شده منجر شده است که این روش‌ها در مقابل حملات مربوطه بخصوص در مواردی که طول پیام کوتاه است امن شوند. توجه شود که نرخ تشخیص ۵۰٪ به معنی آن است که روش نهان‌کاوی مورد نظر قادر به تشخیص تصاویر نهان‌نگاری شده نیست.

**جدول ۱- درصد تشخیص صحیح روش‌های نهان‌کاوی [۳،۴].**

| طول پیام و نوع حمله | LSB±1 قبل از بهبود | LSB±1 بعد از بهبود | LSB±1(R) قبل از بهبود | LSB±1(R) بعد از بهبود |
|---|---|---|---|---|
| 0.2 bpp[3] | 56 | 52 | 51 | 50 |
| 0.2 bpp[4] | 55 | 51 | 50 | 50 |
| 0.4 bpp[3] | 63 | 55 | 55 | 51 |
| 0.4 bpp[4] | 60 | 54 | 57 | 50 |
| 0.6 bpp[3] | 74 | 61 | 63 | 51 |
| 0.6 bpp[4] | 69 | 58 | 62 | 51 |
| 0.8 bpp[3] | 84 | 63 | 68 | 54 |
| 0.8 bpp[4] | 79 | 59 | 65 | 52 |

## ۶- نتیجه‌گیری

در این مقاله دو روش رایج از تکنیک‌های نهان‌نگاری شـامل درج در کم‌ارزش‌ترین بیت تطبیقی و درج درکم‌ارزش‌تـرین بیـت تطبیقـی بازبینی شده بهبود داده شد. این دو روش منجر به تغییرات زیـادی در ماتریس خود همبستگی شده و زمینه تشخیص تصاویر نهان‌نگاری شده را توسط تعدادی از روش‌های نهان‌کاوی فراهم کرده است. با توجه بـه اینکه در فرآیند درج توسط این دو روش امکان کاهش یا افزایش مقدار پیکسل وجود دارد، می‌توان عمل کاهش یا افزایش را بصورتی انجام داد که منجر به تغییرات کمتری در ماتریس خودهمبستگی شود. مـا ایـن بهبود را به دو روش مذکور اعمال کردیم. نتایج آزمایشات نشان می‌دهد که بهبود داده شده منجر شده است که عملیـات تغییـرات کمتـری در ماتریس خودهمبستگی بوجود آورد و این روش‌ها در مقابل حملاتی که از ماتریس خودهمبستگی برای نهان‌کاری بهره می‌برند امن شوند.

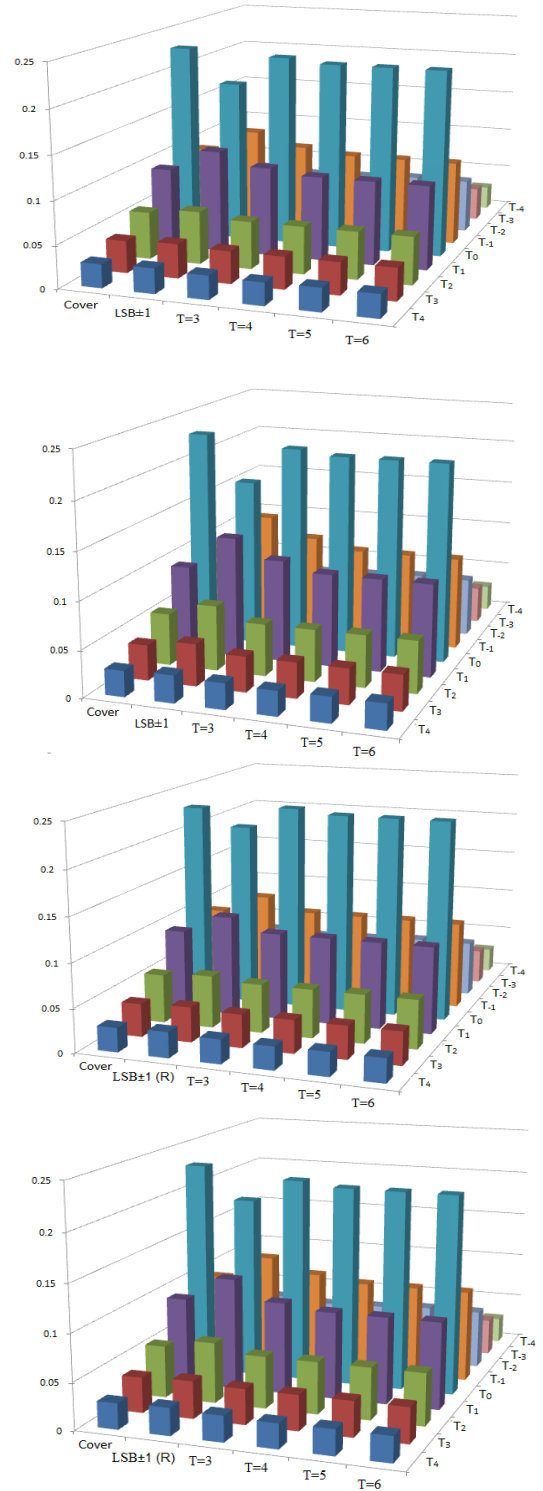

شکل ۴- انرژی قطر اصلی ($T_0$) و قطرهای نزدیک به قطر اصلی ($T_{-4}$ $T_{-3}$ $T_{-2}$ $T_{-1}$ $T_1$ $T_2$ $T_3$ $T_4$) ماتریس خودهمبستگی برای تصویر اصلی (Cover) و تصویر نهان‌نگاری شده با LSB±1 و LSB±1(R) و بهبود یافته آنها با مقادیر مختلف پارامتر T. از بالا به پایین برای طول پیام ۰/۴، ۰/۸، ۰/۴ و ۰/۸. دو تصویر بالاتر برای LSB±1 و دو تصویر پایین برای LSB±1(R).